\documentclass{article}
\usepackage{spconf,amsmath,graphicx}
\usepackage{bm,epsfig}
\usepackage{algorithm}
\usepackage{multirow}
\usepackage{booktabs}
\usepackage{amsmath,amsfonts,amssymb,graphicx}

\makeatletter
\def\hlinewd#1{%
  \noalign{\ifnum0=`}\fi\hrule \@height #1 \futurelet
   \reserved@a\@xhline}
\makeatother
\title{Low Bit-Rate Wideband Speech Coding: A Deep Generative Model based Approach}
\name{Gang Min$^1$, Xiongwei Zhang$^2$, Xia Zou$^2$, Xiangyang Liu$^1$}
\address{$^1$Institute of Information and Communication, National University of Defense Technology, China\\
  $^2$Army Engineering University of PLA, China}

\begin{document}
%
\maketitle
\begin{abstract}
Traditional low bit-rate speech coding approach only handles narrowband speech at 8kHz, which limits further improvements in speech quality. Motivated by recent successful exploration of deep learning methods for image and speech compression, this paper presents a new approach through vector quantization (VQ) of mel-frequency cepstral coefficients (MFCCs) and using a deep generative model called WaveGlow to provide efficient and high-quality speech coding. The coding feature is sorely an 80-dimension MFCCs vector for 16kHz wideband speech, then speech coding at the bit-rate throughout 1000-2000 bit/s could be scalably implemented by applying different VQ schemes for MFCCs vector. This new deep generative network based codec works fast as the WaveGlow model abandons the sample-by-sample autoregressive mechanism. We evaluated this new approach over the multi-speaker TIMIT corpus, and experimental results demonstrate that it provides better speech quality compared with the state-of-the-art classic MELPe codec at lower bit-rate.
\end{abstract}
\begin{keywords}
speech coding, mel-frequency cepstral coefficients, vector quantization, WaveGlow
\end{keywords}
\section{Introduction}
\label{sec:intro}

Low bit-rate speech coding, which encodes speech signals at the bit rate below 4800 bit/s, has widespread applications in the field of both satellite and secure communications. Many successful low bit-rate speech coding algorithms have been proposed in the literatures, such as linear predictive coding (LPC-10) \cite{Tremain:LPC10}, code-excited linear prediction (CELP) \cite{ICASSP:CELP}, mixed excitation linear prediction (MELP) \cite{TSAP:McCree}, etc. However, high-quality speech coding under low bit-rate conditions still faces great challenge, especially for the wideband speech and in the presence of background acoustic noises. All of the classic speech vocoders mentioned above belong to the source-filter speech coding framework, in which the speech coding parameters include linear prediction coefficients (LPCs), pitch, energy, etc. Different types of speech coding parameters are rarely quantized together, so it is very difficult to further reduce the speech coding rate. Therefore, many other speech coding methods have been studied towards alternatives to the classic linear prediction coding model.

MFCC codec encodes speech signals through scalar quantization (SQ) or vector quantization (VQ) of MFCCs, which provides a new promising scheme for speech coding at low bit-rate conditions \cite{TASLPLow:Boucheron}\cite{SPL:Gang}. However, there are still some limitations need to be resolved for further improving the total performance. The first is the quality of coded speech needs further improvement, since there exists spectrum smearing problem, especially in the high-frequency region, which is caused by using the overlapped triangle window with mel-frequency scale for MFCCs extraction. Another is the processing efficiency also needs improvement since the traditional MFCC codec uses the Griffin-Lim algorithm (GLA) to estimate the lost phase information via discrete Fourier transform (DFT) and inverse discrete Fourier transform (IDFT) iteratively \cite{TASSP:Griffin}. However, GLA suffers from slow convergence problem when the random initialization of the phase spectrogram is not ideal. Moreover, current MFCC codec is rarely able to handle 16kHz wideband speech signals \cite{TASLPLow:Boucheron}.

In the last decade, deep learning methods have been used for dramatically improving the performance of many speech processing applications, such as speech enhancement (SE), text-to-speech (TTS), automatic speech recognition (ASR), etc. Most recently, deep neural networks have shown to be promising in handling the traditional speech coding task \cite{Interspeech2019:Backstrom}. One of the most representative works is WaveNet based codec \cite{ICASSP2018:Kleijn}\cite{ICASSP2019:Garbacea}, which uses WaveNet as a generative model to synthesize speech waveforms from the bitstream generated from traditional speech codecs, such as codec2, MELP, etc. WaveNet is a kind of autoregressive neural networks-based model which generates high-quality speech waveforms, however, WaveNet suffers from very slow inference speed, which prevents its real-time speech coding applications. Besides, other models such as simple RNN, LPCNet are also explored for speech and audio coding applications \cite{ICASSP2019:Klejsa}-\cite{ICASSP2020:Fehgin}. The authors in \cite{ICMEW:Gang} presented Deep Vocoder, which compresses narrowband speech with deep autoencoder and uses GLA to recover speech signals from decoded speech spectrogram, similar work in \cite{ACCESS:Keles} presented DeepVoCoder which uses a convolutional neural network (CNN)-based encoder model to compress speech signals. However, the quality of coded speech and the efficiency of speech decoding need further improvement for real-word communication applications.

Recent research on TTS using deep generative models conditioned on mel-spectrogram motivates our study in combining quantization of MFCCs with efficient and high-quality speech generative models for speech coding task in this paper. WaveGlow is a flow-based deep generative network, which delivers speech quality almost as good as WaveNet, however, the inference speed of WaveGlow is much faster than which of WaveNet because it abandons the sample-by-sample autoregressive mechanism \cite{ICASSP2019:Prenger}. Recent study of comparison on neural vocoders for speech reconstruction from mel-spectrogram also confirmed the superiority of WaveGlow for making tradeoff between speech quality and computational complexity \cite{ISCA:Govalkar}. Therefore, we choose WaveGlow as the generative model to synthesize speech waveforms from the quantized mel-spectrogram. The coding feature in our vocoder is an 80-dimension MFCCs vector for 16kHz wideband speech signal, then speech coding at the bit-rate throughout 1000-2000 bit/s could be scalably implemented with different quantization schemes of MFCCs vector.

\section{Algorithm}
\subsection{Speech Coding with MFCCs and WaveGlow}

Speech coding model is the basis for converting speech signals to bitstream. Like traditional speech vocoders, there are mainly three steps for speech coding with quantization of MFCCs and WaveGlow, which are extraction of speech coding features, quantization of these features and speech synthesis from quantized feature parameters, as is shown in Fig.1.
\begin{figure}[htb]
  \centering
  \centerline{\includegraphics[width=8.5cm]{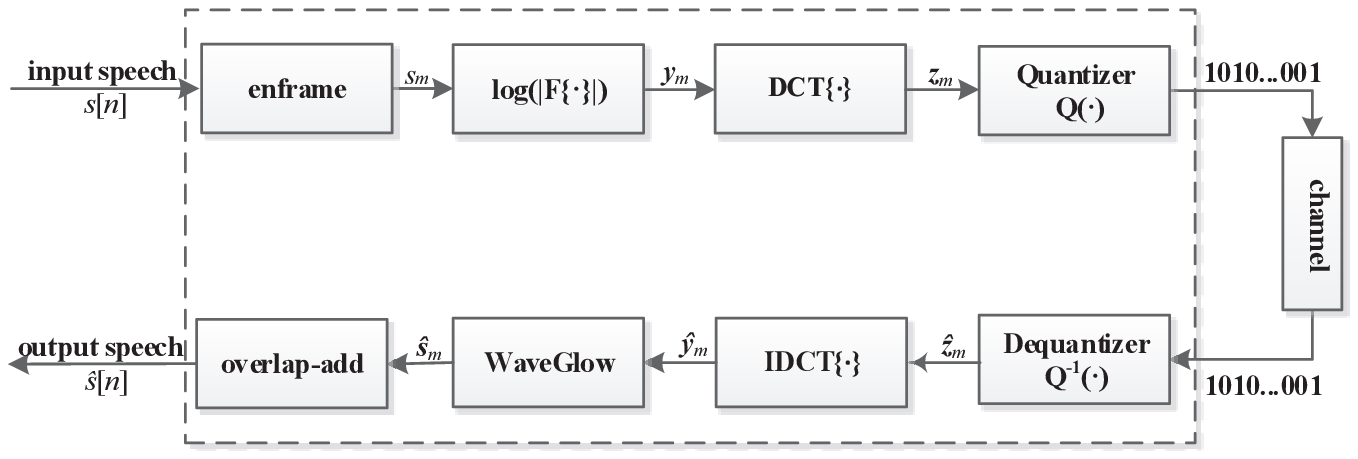}}
  \caption{Overview of the proposed vocoder.}
\end{figure}

Let \emph{s}[\emph{n}] denote the speech waveforms, then it is enframed by a window $w\left[n\right]$,
\begin{equation}
{s_m}\left[n\right]=s\left[{mR+n}\right]w\left[n\right]
\end{equation}
where $L (0\leq{n}\leq{L-1})$ denotes the window length, \emph{R} denotes the frame shift, $m(m=1,2,...,M)$ denotes the frame index. At this time, each speech frame is concisely denoted as follows,
\begin{equation}
{\bm{s}_m}={\left[ {{s_m}\left( 0 \right),{s_m}\left( 1 \right),...,{s_m}\left( {L - 1} \right)} \right]^{\intercal}}
\end{equation}

Then, the log mel-spectrogram of each speech frame can be computed as,
\begin{equation}
{\bm{y}_m}=\mathbf{M}\log(\left|{{\rm{F}}\left\{{\bm{s}_m}\right\}}\right|)
\end{equation}
where ${\rm{F}}\left\{{\bm{s}_m}\right\}$ is the \emph{N}-point fast Fourier transform (FFT) of $\bm{s}_m$, $\left|\cdot\right|$ denotes the modulus of a complex number. Due to the symmetry, the latter $N/2-1$ elements of $\left|{{\rm{F}}\left\{{\bm{s}_m} \right\}}\right|$ will be discarded. $\mathbf{M}\in\mathbb{R}^{K \times (N/2+1)}$ denotes the mel-filter weighting matrix, where $K$ is the number of mel-filter bands.

Furthermore, the MFCCs vector of each speech frame can be computed as follows,
\begin{equation}
{\bm{z}_m}=\mathrm{DCT}\left\{\bm{y}_m\right\}
\end{equation}
where $\mathrm{DCT}\left\{\cdot\right\}$ denotes the discrete cosine transform.

At the transmitter, the quantizer $Q(\cdot)$ uses the SQ or VQ technique to quantize MFCCs vector $\bm{z}_m$, and converts it to bitstream, which is then modulated for transmitting.

At the receiver, the quantized MFCCs vector ${{\bm{\hat z}}_m}$ is recovered by searching the codebook by the dequantizer $Q^{-1}(\cdot)$. Then, the reconstructed log mel-sepctrogram ${{\bm{\hat y}}_m}$ is computed by inverse discrete cosine transform (IDCT) of ${{\bm{\hat z}}_m}$, which is then used for conditioning of WaveGlow in order to synthesize speech frame. At last, the speech waveforms $\hat s[\emph{n}]$ is reconstructed by the overlap-add operation.
\subsection{Quantization of MFCCs}
The quantization step of feature parameters is crucial for reducing the bit-rate of speech coding and maintaining high-quality of coded speech. Conventional speech vocoders contain different types of speech coding parameters, which are rarely quantized together. However, the speech coding parameters in the proposed vocoder mentioned above are solely MFCCs vector, so scalable speech coding schemes at different bit-rate could be implemented conveniently using the SQ or VQ technique. The first element of MFCCs vector represents energy, where its value and variance is significantly greater than other elements, so it is independently quantized using the SQ technique. As for other elements of MFCCs vector, they represent the vocal and excitation parameters, which are quantized together using the VQ technique.
\begin{table*}[htp]
\caption{Bit allocation scheme for MFCCs quantization.}
\centering
{\begin{tabular}{ccccccc}
\hlinewd{0.9pt}
$f_{s}$ &$L$          &$R$         &Rate      & {Bits/}   & \multicolumn{2}{c}{Quantizaiton Scheme}                               \\\cline{6-7}
(Hz)    &{(sample)}   &{(sample)}  &(bit/s)   & frame     & Energy($z_{0}$)                 & Formant and Pitch ($z_{2}\sim z_{80}$)   \\\hline
16000   &1024         &256         &1000      & 16        & \multicolumn{1}{c}{4-bit SQ}    & \multicolumn{1}{c}{12-bit VQ}          \\
16000   &1024         &256         &2000      & 32        & \multicolumn{1}{c}{6-bit SQ}    & \multicolumn{1}{c}{(13-13)-bit MSVQ}     \\
\hlinewd{0.9pt}
\end{tabular}}{}
\end{table*}
\subsection{Conditional WaveGlow as a Decoder}

WaveGlow is a flow-based deep neural generative model for synthesizing high-quality speech signals conditioned on mel-spectrogram. Previous study has shown that Mean Opinion Score (MOS) of the synthesized speech via WaveGlow is able to reach up to 3.9 on the LJ speech corpus \cite{ICASSP2019:Prenger}, so trying to use WaveGlow as a decoder for speech coding is very attractive. WaveGlow consists of a series of invertible flow layers that transforms a simple zero mean spherical Gaussian distribution to one which has the desired speech distribution \cite{ICASSP2019:Prenger}. WaveGlow network could be directly trained by minimizing the negative log-likelihood of for training set. Once the WaveGlow network is trained, doing inference to generate speech waveforms from quantized mel-spectrogram could be implemented by sampling from a Gaussian distribution and putting them through the WaveGlow network.

\subsection{Bit Allocation Scheme for Speech Coding}

Bit allocation is an important procedure for determining the bit-rate of speech coding. As previously discussed, the first element of MFCCs vector $z_{0}$ and other elements of MFCCs vector $z_{2}\sim z_{80}$ are quantized using different methods, respectively. The proposed vocoder proceeds with the wideband speech signals (16kHz sampling rate), when the frame length is set as 64 msec (1024 samples) and the frame shift is set as 16 msec (256 samples), respectively, we can design the bit allocation scheme as is shown in Tab.1. We can see that speech coding at different bit-rates could be flexibly implemented given the corresponding bit allocation schemes.

When the bit-rate is 1000 bit/s, there are totally 16 bits for each speech frame, so only 4 bits are allocated for scalar quantization of energy parameter $z_{0}$ and the last 12 bits are allocated for direct vector quantization of formant and pitch parameters $z_{2}\sim z_{80}$. When the bit-rate is 2000 bit/s, there are totally 32 bits for each speech frame, so 6 bits are allocated for scalar quantization of energy parameter and another 26 bits are allocated for quantization other parameters.

In order to reduce the codebook searching complexity at the bit-rate of 2000 bit/s, we use multistage vector quantization (MSVQ) method to encode $z_{2}\sim z_{80}$ efficiently. To make a tradeoff between the quantizing distortion and codebook searching burden, 2 cascaded codebooks are trained and the codebook at each stage consists of $2^{13}$ codewords, the quantization result of $z_{2}\sim z_{80}$ is computed by comparison on quantization distortion of different combination of the reserved codewords at each stage.

\section{Experiments and Results}
\label{sec:pagestyle}

\subsection{Dataset and Evaluation Metrics}
We carry our experiments on the widely used TIMIT corpus to evaluate the performance of the proposed vocoder. TIMIT is a multi-speaker corpus, which contains 462 speakers in the training dataset and 168 speakers in the test dataset. At the training stage, the whole TIMIT training set with 4620 utterances were used for extracting mel-spectrograms and training the WaveGlow network model, the duration of the training speech is $\thicksim$4 hours. At the test stage, we chosen 300 utterances spoken by a total of 30 speakers from the test dataset for speech coding, the duration of test speech is $\thicksim$16 minutes. All the speech waveforms are sampled at 16kHz. The speech signal was enframed to 1024 samples using a hamming window and the frame shift is 256 samples. The dimension of MFCCs vector for each speech frame is 80, i.e., $K=80$.

Two different objective metrics were used for evaluating the quality of coded speech. The first is perceptual evaluation of speech quality (PESQ) \cite{ICASSP:Rix}, which is adopted as the ITU-T P.862 standard and widely used for evaluating speech quality. Another is is the short-time objective intelligibility (STOI) \cite{TASLPAn:Taal}, which is also a popular objective measure. PESQ demonstrates the overall speech quality while the STOI measure illustrates the speech intelligibility. For both the metrics, higher score indicates better performance. Also, we will take some subjective listening experiments to further demonstrate the performance of the proposed method.
\subsection{Hyper-parameters Setting for WaveGlow Training}
WaveGlow model was usually trained on single-speaker corpus for speech synthesis in previous study. However, speech coding for multi-speakers is much usual in real-word communication applications. Therefore, to obtain a good multi-speaker WaveGlow model on TIMIT corpus, the hyper-parameters should be carefully configured. Considering both the performance of WaveGlow network and the capacity of our hardware platform (Intel Xeon CPU (2.2GHz), 128G RAM and NVIDIA GeForce GTX 1080Ti $\times2$ GPUs), we configured the hyper-parameters of WaveGlow as is shown in Tab.2. The quantized and unquantized mel-spectrogram were independently used as the input for WaveGlow training, the ADAM algorithm was chosen as the optimizer with the learning rate as $1\times10^{-4}$. After 1,110,000 epoches of training, we obtained a WaveGlow network model which was used as a decoder for low bit-rate speech coding.
\begin{table}[htb]
  \caption{hyper-parameters setting for WaveGlow training}
  \label{tab:hpSettings}
  \centering
  \begin{tabular}{ll}
    \toprule
    hyper-parameter                               & value                  \\
    \midrule
    number of flows                                & 12                    \\
    number of mel-channels                         & 80                    \\
    number of groups                               & 8                     \\
    number of layers for coupling module        & 8                     \\
    number of mel-channels for coupling module  & 256                   \\
    kernel size for coupling module             & 3                     \\
    learning rate                                  & $1\times10^{-4}$      \\
    batch size                                     & 12                    \\
    \bottomrule
  \end{tabular}
\end{table}

\begin{figure*}[htb]
\centering
\includegraphics[width=6in]{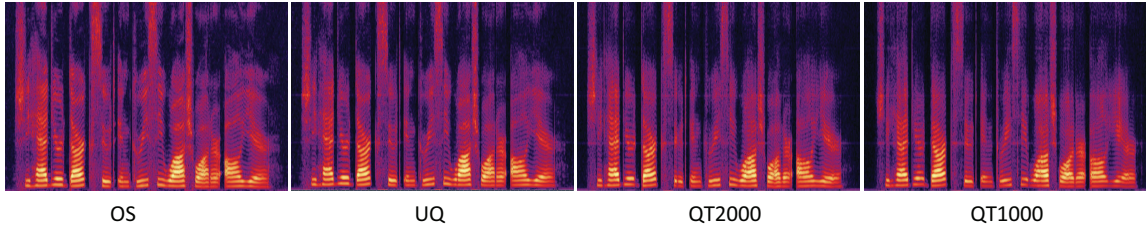}
\caption{Comparison on spectrograms of the TIMIT utterance ``She had your dark suit and greasy wash water all year''.}
\label{fig_deepvocoder}
\end{figure*}

\subsection{Evaluation of speech quality}

For simplicity, we denote the proposed speech coding algorithm via quantization of MFCCs and WaveGlow as WaveGlow codec, some other notations are as follows,\\
$\bullet$ OS: original speech signal\\
$\bullet$ UQ: speech synthesis from unquantized MFCCs\\
$\bullet$ UQT2000: WaveGlow codec at 2000 bit/s with unquantized MFCCs as input for training WaveGlow model\\
$\bullet$ UQT1000: WaveGlow codec at 1000 bit/s with unquantized MFCCs as input for training WaveGlow model\\
$\bullet$ QT2000: WaveGlow codec at 2000 bit/s with quantized MFCCs as input for training WaveGlow model\\
$\bullet$ QT1000: WaveGlow codec at 1000 bit/s with quantized MFCCs as input for training WaveGlow model\\

\begin{figure}[htb]
  \centering
  \centerline{\includegraphics[width=6cm,height=3cm]{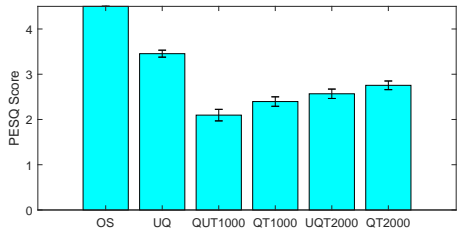}}
  \caption{speech quality in terms of PESQ score.}
\end{figure}

\begin{figure}[htb]
  \centering
  \centerline{\includegraphics[width=6cm,height=3cm]{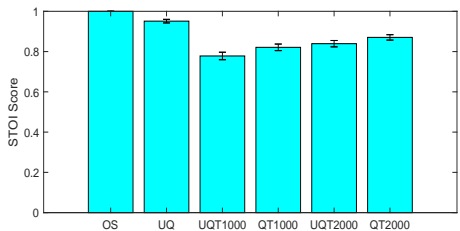}}
  \caption{speech quality in terms of STOI score.}
\end{figure}
Fig.2 shows the spectrograms of the reconstructed speech via WaveGlow codec for a typical TIMIT utterance. We can see that the structure of harmonic and frequency formant is both well preserved, which demonstrates that the original speech and the coded speech sounds closely. Fig.3 and Fig.4 shows the speech quality in terms of PESQ and STOI scores for the test set. It should be noted that WaveGlow trained with quantized MFCCs performs better than WaveGlow trained with unquantized MFCCs, because it overcomes the dismatch problem during the WaveGlow training and inference stage. We can also see that the output speech quality for QT2000 and QT1000 is acceptable as the PESQ scores of the output speech is about 2.75 and 2.52, respectively. We listened these coded speech signals and found that the output speech of WaveGlow codec preserves high intelligibility and somewhat naturalness though few audible artifacts exist.

We also conducted subjective listening tests. 10 volunteers rated the coded speech through the standard five point mean opinion score (MOS) \cite{MOS:Loizou}. Each volunteer was presented with 20 speech files encoded by WaveGlow codec and MELPe codec. The results are illustrated in Fig.5, which illustrates that WaveGlow codec provide substantially improved speech quality than MELPe codec at similar bit-rate. In detail, the MOS score for QT2000 and QT1000 is about 3.25 and 2.96, respectively.
\begin{figure}[htb]
  \centering
  \centerline{\includegraphics[width=6cm]{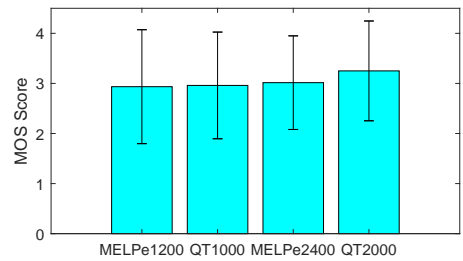}}
  \caption{speech quality in terms of MOS score.}
\end{figure}

\section{Conclusions}

This paper presented a new low bit-rate wideband speech coding approach though vector quantization of MFCCs. WaveGlow was used as a decoder in order to provide efficient and high-quality speech coding at 1000-2000 bit/s. Experimental results demonstrate that WaveGlow codec is promising for low bit-rate source coding of speech signals with high speed inference. In further, other efficient generative models conditioned on mel-spectrogram, such as generative adversarial networks (GANs) \cite{arXiv:Goodfellow}\cite{NIPS2019:Kumar}, are also worth being explored for speech coding purpose. Moreover, the post-filtering technique is also worth studying to reduce the audible artifacts.
\section{Acknowledge}
This work is partially supported by Natural Science Foundation of China(61701535, 61871471) and Key Research and Development Project of Shannxi Province (2020GY-015).

\bibliographystyle{IEEEbib}
\bibliography{strings,refs}

\end{document}